\documentclass[a4paper]{article}
\pdfoutput=1

\usepackage{jheppub}
\usepackage{macros}
\usepackage{physics}
\usepackage{makecell}

\preprint{ }

\title{A note on half-integer irregular representations of Virasoro algebra}

\author[1]{Yichi Zang}

\affiliation[1]{Department of Physics and Center for Field Theory and Particle Physics, Fudan University, \\
20005, Songhu Road, 200438 Shanghai, China}

\emailAdd{yichyzang@gmail.com}

\abstract{
We study irregular representations of Virasoro algebra associated with half-integer order singularities, which arise naturally in the 2d CFT description of Argyres-Douglas theories of type $(A_1, A_{\text{even}})$ and $(A_1, D_{\text{odd}})$. While integer-rank irregular states admit a well-established free-field construction, the half-integer case is more subtle due to the presence of branch cuts. In this note, we present two equivalent constructions of half-integer irregular representations. The first one is based on a $\bZ_2$-twisted free boson, which is motivated from the monodromy structure of Hitchin system. The second one employs a recursion relation of the Virasoro eigenvalues recently proposed in the literature. We explicitly demonstrate the equivalence of these two parameterization schemes at rank $3/2$ and $5/2$. Our analysis clarifies the structure of half-integer irregular modules and provides tools for computing the corresponding irregular states relevant for Argyres-Douglas theories.
}

\begin{document}
\allowdisplaybreaks
\setcounter{tocdepth}{3}
\maketitle

\section{Introduction}

The discovery of the Alday-Gaiotto-Tachikawa (AGT) correspondence has unveiled profound connections between 4d $\mathcal{N}=2$ supersymmetric gauge theories and 2d conformal field theories (CFTs) \cite{Alday:2009aq,Wyllard:2009hg}. In its simplest form, the AGT relation identifies the Nekrasov partition function of certain 4d $\mathcal{N}=2$ theories, particularly those arising from class $\mathcal{S}$ constructions \cite{Gaiotto:2009we}, with conformal blocks of 2d CFTs such as Liouville or Toda theories. This duality has provided deep insights into both supersymmetric gauge dynamics and the representation theory of infinite-dimensional algebras.

A particularly intriguing subclass of 4d $\mathcal{N}=2$ theories are the so-called Argyres-Douglas (AD) theories \cite{Argyres:1995jj,Argyres:1995xn}. These theories, first discovered through the study of special singularities on the Coulomb branch of conventional gauge theories, exhibit several exotic features: they are strongly interacting, admit no known Lagrangian description, and possess Coulomb branch operators with fractional scaling dimensions. AD theories have since been realized systematically within the framework of class $\mathcal{S}$, where they arise from compactifications of the 6d $\cN=(2,0)$ theory on a Riemann sphere with an irregular (higher-order) puncture and potentially one more regular puncture in the associated Hitchin system \cite{Xie:2012hs,Wang:2015mra}.

In the context of the AGT correspondence, the presence of irregular singularities modifies the structure of the associated 2d CFT. Rather than ordinary conformal blocks built from primary fields, the relevant objects are irregular conformal blocks, constructed from coherent states (sometimes called Whittaker states) that satisfy relaxed highest weight conditions in the Virasoro or $W$-algebra modules. These irregular blocks capture the nontrivial asymptotic behavior induced by the irregular singularities and encode the instanton contributions of the corresponding 4d theory \cite{Gaiotto:2009ma,Bonelli:2011aa,Felinska:2011tn,Gaiotto:2012sf,Kanno:2013vi,Nishinaka:2019nuy,Kimura:2020krd,Poghosyan:2023zvy,Hamachika:2024efr,Poghossian:2025nef,Iorgov:2025hxt}.

In this short note, we turn to the Virasoro algebra and study irregular states of half-integer rank. In the rest of the note, we first briefly review the integer rank irregular states and the free-field method used in \cite{Gaiotto:2012sf} to realize this representation. Then we introduce two different ways to characterize the half-integer rank irregular states. In subsection \ref{sec:half-integer}, we explain a modified free-field representation which uses a $\bZ_2$-twisted free scalar field to realize the irregular Virasoro action of half-integer rank. Then in subsection \ref{sec:Lambda_rep}, we show an alternative way of parameterize the state directly using eigenvalues of certain Virasoro generators, as proposed in \cite{Hamachika:2024efr}. In their paper, they introduce a recursive way to determine to complete representation. We follow their method and provide a closed-form solution to their recursion. This result can be used to directly compute the representation using combinatorial methods. Finally, we use two examples to demonstrate that these two parameterizations are equivalent, as they can be related by a suitable change of variables and normalization. A similar argument should hold for generic rank even though we only provide the explicit construction at rank $3/2$ and $5/2$.

\section{Half-integer irregular representations for Virasoro algebra}

While the construction of integer-rank irregular states has been extensively discussed in the literature, the half-integer case is more subtle and requires a slightly different treatment. These states naturally arise, for instance, in the context of Argyres–Douglas theories and in the study of irregular singularities of type $z^{-m-\tfrac{1}{2}}$ in two-dimensional conformal field theory.

The class $\cS$ construction of AD theories has been given in \cite{Xie:2012hs}. AD theories of type $(A_1,A_{2n-3})$ are associated to Hitchin systems on a sphere with an irregular puncture where the Higgs field has a higher-order pole at zero
\be \label{Higgs-pole-2}
\Phi=\lambda_1\left(\begin{array}{cc}
1 & 0 \\
0 & -1
\end{array}\right) \frac{\dd{z}}{z^{n+1}}+\cdots
\ee 
where $n\in \bN/2$. When $n$ is a strictly positive integer, 
the corresponding irregular state $|I^{(n)}\rangle$ of Virasoro algebra was characterized in \cite{Gaiotto:2012sf} as a coherent state on which $\qty{L_n,\dots,L_{2n}}$ are simultaneously diagonalized
\be
L_k \ket*{I^{(n)}} = \Lambda_k \ket*{I^{(n)}} \qfor n\leqslant k \leqslant 2n.
\ee
From the commutation relation of Virasoro algebra, we can infer that higher generators $\qty{L_{k>2n}}$ must annihilate the irregular state $\ket*{I^{(n)}}$, and $\qty{L_0,\dots,L_{n-1}}$ act on the state as differential operators.
Using the free-field representation from the stress-energy tensor
\begin{equation}\label{eq:Liouville_stress_energy}
T(z)=-: \partial_z \phi(z) \partial_z \phi(z):+Q \partial_z^2 \phi(z),
\end{equation}
we can construct the explicit Virasoro action on the state by considering a coherent state of the lowering modes $\qty{\alpha_0,\dots,\alpha_n}$ of $\partial\phi(z)$.
For a positive integer $n$, the irregular state $|I^{(n)}\rangle$ satisfies the following set of differential equations:
\be
L_k |I^{(n)}\rangle = 
\left\{
\begin{aligned}
&0 && \text{for} \quad k > 2n, \\[2mm]
&\Lambda_k |I^{(n)}\rangle && \text{for} \quad n \leq k \leq 2n, \\[2mm]
&\left(\Lambda_k + \sum_{\ell=1}^{n-k} \ell c_{\ell+k} \frac{\partial}{\partial c_\ell}\right) |I^{(n)}\rangle && \text{for} \quad 0 \leq k \leq n-1,
\end{aligned}
\right.
\label{eq:diff-eq-int}
\ee
where $\qty{c_0,\dots,c_n}$ are eigenvalues of corresponding $\alpha_k$, and the coefficients $\qty{\Lambda_0, \dots, \Lambda_{2n}}$ can be expressed by $\qty{c_0, \dots, c_n}$ according to
\be
\Lambda_k \equiv 
\left\{
\begin{aligned}
&-\sum_{i=k-n}^{n} c_i c_{k-i} && \text{for} \quad n < k \leq 2n, \\[2mm]
&-\sum_{i=0}^{k} c_i c_{k-i} + (k+1)Q c_k && \text{for} \quad 0 \leq k \leq n.
\end{aligned}
\right.
\ee
It was further shown in \cite{Gaiotto:2012sf,Nishinaka:2019nuy} that, when $n$ is an integer, a power series expansion for $|I^{(n)}\rangle$ can be systematically constructed by solving \eqref{eq:diff-eq-int} order by order in $c_n$. The inner product $\langle 0|I^{(n)}\rangle$ corresponds to the instanton partition function of $(A_1,A_{2n-3})$ AD theory. For $(A_1,D_{2n})$ type AD theory, where the associated sphere has one irregular puncture and one regular puncture, the instanton partition function corresponds to the inner product $\langle \Delta |I^{(n)}\rangle$. These inner products have been evaluated in \cite{Nishinaka:2019nuy} for $n=2,3$. 

In contrast, when the degree $n$ of the pole in \eqref{Higgs-pole-2} is a strictly half-integer, 
it is more challenging to construct the corresponding Virasoro irregular states. Nevertheless, such irregular representations have been recently studied  in \cite{Poghosyan:2023zvy,Hamachika:2024efr,Poghossian:2025nef}, and the inner products have been evaluated for $n=\frac{3}{2},\frac{5}{2}$.
Based on these recent developments, in this section, we shall construct irregular representations of Virasoro algebra for a strictly half-integer $n$.

\subsection{Free field representation of half-integer rank}\label{sec:half-integer}

The irregular representations in the half-integer case are obtained by extending the free-field construction \eqref{eq:diff-eq-int} in \cite{Iorgov:2025hxt}.\footnote{The author had independently formulated the same representation prior to the appearance of \cite{Iorgov:2025hxt}; see \cite{Zang_Bachelor}. The construction in \cite{Iorgov:2025hxt}, however, provides a more complete account, and in this work we supplement it by offering a physical interpretation of the resulting structure.}
Conceptually, the half-integer generalization requires adapting the free-field framework to incorporate the branch-cut behavior characteristic of half-integer poles. This modification yields a consistent analogue of the usual free-field representation of irregular states.

A half-integer pole ($n\to n-\frac12\in\bN+\frac12$ in \eqref{Higgs-pole-2}) corresponds to an irregular state of half-integer rank $\ket*{I^{(n-\frac12)}}$, which is defined as a coherent state of $\qty{L_n,\dots,L_{2n-1}}$, and is annihilated by $\qty{L_{k\geqslant 2n}}$.
The naive application of the free-field representation of the stress-energy tensor \eqref{eq:Liouville_stress_energy} as in \eqref{eq:diff-eq-int} for this half-integer pole does not produce desired representation.
Nevertheless, inspired by the branch cutting behavior of a half-integer pole, we construct an analogue of \eqref{eq:diff-eq-int} with a $\bZ_2$-twisted field \cite{dixon:1986qv}.

Consider a $\bZ_2$-twist line defect from $z=0$ to $z=\infty$, with two twist operators placed at endpoints $\sigma(0)$ and $\sigma(\infty)$. A free chiral scalar field in the twisted sector exhibits the monodromy behavior
\begin{equation}
    \phi(z)\sigma(0) \sim z^{1/2}\tau(0).
\end{equation}
The mode expansion of $\partial\phi(z)$ is given by
\begin{equation}\label{eq:mode_twisted_field}
    \partial\phi(z) = -i\sum_{r\in\bZ+\frac12} \tilde{\a}_r z^{-r-1},
\end{equation}
with commutation relation
\begin{equation}
    \comm*{\tilde{\a}_r}{\tilde{\a}_s} = \frac{r}{2}\delta_{r+s,0}.
\end{equation}
There is an issue that stress-energy tensor in \eqref{eq:Liouville_stress_energy} does not admit the $\bZ_2$ symmetry. Hence we need to work with a new stress-energy tensor
\begin{equation}
    T(z) = -:\partial\phi(z)\partial\phi(z): \, ,
\end{equation}
which is the $Q=0$ limit of \eqref{eq:Liouville_stress_energy}. In this case, the Virasoro generators are given by
\begin{equation}\label{eq:Virasoro_mode_half_int}
    L_k = \sum_{r\in\bZ+\frac12} :\tilde{\a}_{k-r} \tilde{\a}_r : .
\end{equation}
To construct the analogue of free field representation, we can consider a coherent state that
\begin{equation}
    \tilde{\a}_{r} \ket*{{\pmb{b}}^{(n-\frac12)}} = \left\{
    \begin{aligned}
        &0 && r \geqslant n+\frac12,\\
        &-ib_{r} \ket*{{\pmb{b}}^{(n-\frac12)}} && \frac12\le r\le n-\frac12,\\
        &\frac{i}{2}(-r)\pdv{b_{-r}} \ket*{{\pmb{b}}^{(n-\frac12)}} && -n+\frac12 \le r\le-\frac12.
    \end{aligned}
    \right.
\end{equation}
Such a state can be used to construct an irregular representation of half-integer rank $\irreg{n-\frac12}$. The explicit form of Virasoro generator actions is
\begin{equation}\label{eq:irregular_c_half_int}
    L_k\irreg{n-\frac12} = \left\{
    \begin{aligned}
        &0 && 2n \le k, \\
        &-\sum_{r=k-n+\frac12}^{n-\frac12} b_{k-r}b_{r} \irreg{n-\frac12} && n\le k<2n, \\
        &\qty( -\sum_{r=\frac12}^{k-\frac12}b_{k-r}b_{r} + \sum_{r=\frac12}^{n-k-\frac12}r b_{r+k} \pdv{b_{r}}) \irreg{n-\frac12} && 0\le k< n.
    \end{aligned}
    \right.
\end{equation}
In this construction, the $L_0$ action only contains differential operator. We will see in section \ref{sec:ex} that this is aligned with the result given in the recent research where they also find $f_0=0$ in \eqref{eq:half-int_lambda} \cite{Poghosyan:2023zvy,Hamachika:2024efr}.

\subsection{Another representation}\label{sec:Lambda_rep}

There exists yet another approach to the Virasoro representations of half-integer irregular states \cite{Hamachika:2024efr}. 
In this subsection, we present the general structure of such states and clarify how the parameters entering the Virasoro action are organized.

In \cite{Hamachika:2024efr}, it is proposed that a rank-\((n-\tfrac12)\)
irregular state can be parametrized by the set
\(\{\Lambda_{k}\}_{k=n}^{2n-1}\).  
The action of the Virasoro generators on such a state is given by
\begin{equation}\label{eq:half-int_lambda}
    L_k\irreg{n-\frac12} = \left\{
    \begin{aligned}
        &0 && 2n \le k,\\
        &\Lambda_k\irreg{n-\frac12} && n\le k<2n,\\
        &\qty(f_k(\Lambda) + \sum_{\ell=n}^{2n-k-1}(\ell-k) \Lambda_{\ell+k}\pdv{\Lambda_{\ell}}) \irreg{n-\frac12} && 0\le k<n,
    \end{aligned}
    \right.
\end{equation}
where the nontrivial dynamical information is encoded in the functions
\(f_k(\Lambda)\).
The functions \(f_k(\Lambda)\) are conjectured to satisfy an iterative recursion
relation of the form \cite{Hamachika:2024efr}
\be \label{recursion}
\begin{aligned}
f_k(\Lambda)= & \sum_{\ell=1}^{2 n-k-2}(-1)^{\ell-1} \frac{(\ell+n-k-1)!}{(n-k-1)!} \frac{g_{\ell+1}^{(2 n-1) \ell+k}(\Lambda)}{\left(\Lambda_{2 n-1}\right)^{\ell}} \cr
& \quad+\sum_{m=1}^{n-k-1}\left(\sum_{\ell=1}^m(-1)^{\ell-1} \frac{(\ell+n-k-1)!}{(n-k-1)!} \frac{g_{\ell}^{(2 n-1) \ell-m}(\Lambda)}{\left(\Lambda_{2 n-1}\right)^{\ell}}\right) f_{k+m}(\Lambda)
\end{aligned}
\ee 
with
\be 
g_m^i(\Lambda) \equiv \frac{1}{2 \pi i} \oint_{|z|=1} \frac{d z}{z^{1+i}} \frac{1}{m!}\left(\sum_{\ell=n}^{2 n-2} z^{\ell} \Lambda_{\ell}\right)^m .
\ee 
Both \(f_k\) and \(g_m^i\) depend implicitly on the rank parameter \(n\).
An immediate consequence of the recursion \eqref{recursion} is an explicit
formula for the highest nontrivial component \(f_{n-1}\):
\begin{equation}\label{eq:f-n-1}
f_{n-1}(\Lambda)
=
\sum_{\ell=1}^{n-1}
(-1)^{\ell-1}\,
\ell!\,
\frac{
  g_{\ell+1}^{(2n-1)(\ell+n-1)}(\Lambda)
}{
  \Lambda_{2n-1}^{\,\ell}
}.
\end{equation}
This expression serves as the initial condition for recursively determining all
lower \(f_k\), and therefore plays a central role in constructing the full
half-integer irregular representation.

To solve the recursion relation \eqref{recursion} and obtain a closed-form
expression for the functions \(f_k(\Lambda)\), it is convenient to introduce a
combinatorial parametrization based on Young diagrams.  
Let \(\lambda\) be a Young diagram whose parts are strictly less than \(n\),
\begin{equation}
\lambda = (\lambda_1,\lambda_2,\ldots,\lambda_{m+1})
       = \bigl( (n-1)^{a_{n-1}},\ldots,2^{a_2},1^{a_1} \bigr),
\end{equation}
where \(a_i\) denotes the multiplicity of the part \(i\) in~\(\lambda\).  
The size and the length of \(\lambda\) are then
\begin{equation}
|\lambda|
  = \sum_{i=1}^{\ell(\lambda)} \lambda_i
  = n+p,
\qquad
\ell(\lambda)
  = \sum_{i=1}^{n-1} a_i
  = m+1.
\end{equation}
The number of distinct permutations of the entries of~\(\lambda\) is
\begin{equation}
A_\lambda
  = \frac{\ell(\lambda)!}{\prod_i a_i!}.
\end{equation}

Motivated by the structure of the recursion relation, the function
\(f_k(\Lambda)\) is expected to admit an expansion of the form
\begin{equation}
f_k(\Lambda)
  = f_{n-1-p}(\Lambda)
  = \sum_{|\lambda|=n+p}
    \eta_\lambda^{(p,m)}
    \frac{\Lambda_{2n-1-\lambda}}{\Lambda_{2n-1}^{\,m}},
\end{equation}
where the coefficients \(\eta_\lambda^{(p,m)}\) are determined recursively, and
\(\Lambda_{2n-1-\lambda}\) denotes the product
\begin{equation}
\Lambda_{2n-1-\lambda}
   \equiv \prod_{i=1}^{m+1} \Lambda_{2n-1-\lambda_i}.
\end{equation}

Treating $\lambda = (\lambda_1,\lambda_2,\ldots,\lambda_{m+1})$ as an ordered set, we denote by 
$\lambda' \subsetneq \lambda$ an ordered subset, and write 
$\lambda - \lambda' := \lambda \setminus \lambda'$.
According to the recursion relation \eqref{recursion}, the coefficients 
$\eta_{\lambda}^{(p,m)}$ satisfy
\be \label{eta-1}
\eta_\lambda^{(p, m)}=\frac{(-1)^{m-1}}{m+1} \binom{m+p}{p} A_\lambda+\sum_{\substack{
\lambda^{\prime}\subsetneq\lambda\cr
1\le\ell(\lambda-\lambda’)\le p}}(-1)^{\ell(\lambda-\lambda^{\prime})-1} \binom{\ell(\lambda-\lambda^{\prime})+p}{p} A_{\lambda-\lambda^{\prime}} \eta_{\lambda^{\prime}}^{\left(p^{\prime}, m^{\prime}\right)}.
\ee 
To simplify the structure of this recursion, we introduce the shorthand notation
\be
\begin{aligned}
F(\lambda,\lambda')&=(-1)^{\ell(\lambda-\lambda')-1}\binom{\ell(\lambda-\lambda')+|\lambda|-n}{|\lambda|-n}A_{\lambda-\lambda'}\cr
F(\lambda)&=(-1)^{\ell(\lambda)}\binom{\ell(\lambda)-1+|\lambda|-n}{|\lambda|-n}A_\lambda
\end{aligned}
\ee
In terms of these functions, the recursion \eqref{eta-1} can be reorganized as
\be
\begin{aligned}
\eta_\lambda^{(p,m)}&=\frac{F(\lambda)}{m+1}+\sum_{\lambda'\subsetneq\lambda}F(\lambda,\lambda')\eta_{\lambda'}^{(p',m')}\cr
&=\frac{F(\lambda)}{m+1}+\sum_{\lambda'\subsetneq\lambda}F(\lambda,\lambda')\left(\frac{F(\lambda')}{m'+1}+\sum_{\lambda''\subsetneq\lambda'}F(\lambda',\lambda'')\eta_{\lambda''}^{(p'',m'')}\right)\cr
&=\frac{F(\lambda)}{m+1}+\sum_{\lambda'\subsetneq\lambda}\frac{F(\lambda,\lambda')F(\lambda')}{m'+1}+\sum_{\lambda''\subsetneq\lambda'\subsetneq\lambda}\frac{F(\lambda,\lambda')F(\lambda',\lambda'')F(\lambda'')}{m''+1}+\cdots
\end{aligned}
\ee

This iterated expression naturally suggests viewing the recursion in terms of 
chains of nested subsets of $\lambda=\left(\lambda_1, \lambda_2, \cdots, \lambda_{m+1}\right)$. 
Let
\begin{equation}\label{chain}
\lambda = \lambda^{(0)} \supsetneq \lambda^{(1)} \supsetneq \lambda^{(2)} 
\supsetneq \cdots \supsetneq \lambda^{(N)}, 
\qquad 
0 \le N \le p,
\qquad 
|\lambda^{(N)}| \ge n.
\end{equation}
Summing over all such chains yields the compact expression
\begin{equation}\label{eta-2}
\begin{aligned}
\eta_\lambda^{(p, m)}&=\sum_{N=0}^p\sum_{\text{possible chains}}\frac{F(\lambda^{(N)})}{m^{(N)}+1}\prod_{i=0}^{N-1}F(\lambda^{(i)},\lambda^{(i+1)})\cr
&=\sum_{N=0}^p\sum_{\text{possible chains}} \frac{(-1)^{m-N-1}}{m^{(N)}+1} \prod_{i=0}^N \binom{m^{(i)}+p^{(i)}}{p^{(i)}} A_{\lambda^{(i)}-\lambda^{(i+1)}}
\end{aligned}
\end{equation}
Here the parameters along the chain are defined by
\begin{equation}
m^{(i)}=
\begin{cases}
\ell(\lambda^{(N)}) - 1, & i = N, \\[2pt]
\ell(\lambda^{(i)} - \lambda^{(i+1)}), & \text{otherwise},
\end{cases}
\qquad
p^{(i)} = |\lambda^{(i)}| - n.
\end{equation}
This reformulation highlights the combinatorial structure underlying the
recursion and exhibits $\eta_{\lambda}^{(p,m)}$ as a sum over all possible
nested decompositions of~$\lambda$.

In an equivalent form, the coefficients $\eta_{\lambda}^{(p,m)}$ can be rewritten so as to make
their combinatorial structure more transparent. Starting from the chain
$\lambda^{(0)} \supsetneq \lambda^{(1)} \supsetneq \cdots \supsetneq \lambda^{(N)}$,
we obtain
\be 
\begin{aligned}
\eta_\lambda^{(p, m)} & =(-1)^{m-1} \sum_{\text{possible chains}}(-1)^N \prod_{i=0}^N \frac{\left(m^{(i)}+p^{(i)}\right)!}{\left(m^{(i)}\right)!\left(p^{(i)}\right)!} \frac{\left(m^{(i)}\right)!}{\prod_j\left(a_j^{(i)}-a_j^{(i+1)}\right)!} \cr
& =(-1)^{m-1}\sum_{\text{possible chains}} (-1)^N \prod_{i=0}^N \frac{\left(m^{(i)}+p^{(i)}\right)!}{\left(p^{(i)}\right)!\prod_j\left(a_j^{(i)}-a_j^{(i+1)}\right)!} \cr
& =(-1)^{m-1} \sum_{\text{possible chains}} \frac{(-1)^N}{m^{(N)}+p^{(N)}+1} \prod_{i=0}^N A_{Z^{(i)}}
\end{aligned}
\ee 
where $Z^{(i)}$ is a generalized partition built from $\lambda^{(i)}$ as
\be
Z^{(i)} \equiv \qty(\lambda^{(i)}-\lambda^{(i+1)}) \cup (0^{p^{(i)}})
\ee
whose length and size satisfy
\begin{equation}
\ell(Z^{(i)}) = m^{(i)} + p^{(i)},
\qquad
\lvert Z^{(i)} \rvert = p^{(i)} - p^{(i+1)}.
\end{equation}

This formulation emphasizes that each step in the nested chain contributes
a factor $A_{Z^{(i)}}$, encoding the multiplicities arising from the
decomposition $\lambda^{(i)} \setminus \lambda^{(i+1)}$.  
The final factor $(m^{(N)} + p^{(N)} + 1)^{-1}$ reflects the terminal
constraint of the recursion and ensures that $\eta_{\lambda}^{(p,m)}$ is
assembled as a weighted sum over all possible such decompositions.

This formula reproduces the results for $f_i(\Lambda)$ in Appendix A of \cite{Hamachika:2024efr}, and generalizes it to a closed-form expression.


\subsection{Examples}\label{sec:ex}

As we demonstrate above, there are two different ways to construct half-integer irregular representation. These two parameterizations are in fact equivalent to each other. In this subsection, we use two examples of rank-$\frac32$ and rank-$\frac52$ to show this equivalence.

\paragraph{The case of rank-$\frac{3}{2}$}

The free-field representation \eqref{eq:irregular_c_half_int} at rank-$\frac{3}{2}$ is given by
\be
\begin{aligned}
L_3\ket*{I^{(3/2)}}&=-b_{3/2}^2\ket*{I^{(3/2)}}\cr
L_2\ket*{I^{(3/2)}}&=-2b_{1/2}b_{3/2}\ket*{I^{(3/2)}}\cr
L_1\ket*{I^{(3/2)}}&=\qty(-b_{1/2}^2+\frac12b_{3/2}\pdv{b_{1/2}})\ket*{I^{(3/2)}}\cr
L_0\ket*{I^{(3/2)}}&=\qty(\frac12b_{1/2}\pdv{b_{1/2}}+\frac32b_{3/2}\pdv{b_{3/2}})\ket*{I^{(3/2)}}.
\end{aligned}
\ee
As expected, the irregular state $\ket*{I^{(3/2)}}$ is a common eigenstate of $L_2$ and $L_3$, with eigenvalues $-2b_{1/2}b_{3/2}$ and $-b_{3/2}^2$ respectively.

To transform into the $\Lambda$ parameterization, recall that $\Lambda_k$ is defined as the eigenvalue of the corresponding Virasoro generator $L_k$. Therefore, we can perform the change of variables
\be 
\Lambda_3 = -b_{3/2}^2 ~, \qquad \Lambda_2 = -2b_{3/2}b_{1/2},
\ee
and obtain the following representation of Virasoro algebra
\be
\begin{aligned}
L_3\ket*{I^{(3/2)}}&=\Lambda_3\ket*{I^{(3/2)}}\cr
L_2\ket*{I^{(3/2)}}&=\Lambda_2\ket*{I^{(3/2)}}\cr
L_1\ket*{I^{(3/2)}}&=\qty(\frac{\Lambda_2^2}{4\Lambda_3}+\Lambda_3\pdv{\Lambda_2})\ket*{I^{(3/2)}}\cr
L_0\ket*{I^{(3/2)}}&=\qty(2\Lambda_2\pdv{\Lambda_2}+3\Lambda_3\pdv{\Lambda_3})\ket*{I^{(3/2)}}.
\end{aligned}
\ee
This is almost the same as the $\Lambda$ parameterization proposed in \eqref{eq:half-int_lambda}. The only difference is that according the the recursion relation \eqref{recursion}, $f_1(\Lambda)$ should be
\be
f_1(\Lambda) = \frac{\Lambda_2^2}{2\Lambda_3}
\ee
at rank-$\frac{3}{2}$. This problem can be easily fixed by introducing a normalization to the state as
\begin{equation}
|I^{(3/2)}\rangle \longrightarrow \exp(\frac{\Lambda_2^3}{12 \Lambda_3^2})|I^{(3/2)}\rangle.
\end{equation}
In fact, as pointed out by \cite{Hamachika:2024efr}, using different normalization convention could lead to slightly different irregular representations that are essentially equivalent. Therefore, we can see that the free-field representation using $b_k$ as parameters is indeed equivalent to the representation using $\Lambda_k$ parameters up to a normalization. This is the same case as in integer ranks.

\paragraph{The case of rank-$\frac{5}{2}$}

At rank-$\frac{5}{2}$, the representation has more parameters. The free-field representation is given by
\be 
\begin{aligned}
L_5\ket*{I^{(5/2)}}&= -b_{5/2}^2 \ket*{I^{(5/2)}}\cr
L_4\ket*{I^{(5/2)}}&= -2 b_{3/2} b_{5/2} \ket*{I^{(5/2)}}\cr
L_3\ket*{I^{(5/2)}}&= \qty(-b_{3/2}^2-2 b_{1/2} b_{3/2}) \ket*{I^{(5/2)}}\cr
L_2\ket*{I^{(5/2)}}&=\qty( -2 b_{1/2} b_{3/2} +\frac{1}{2} b_{5/2}\pdv{b_{1/2}} )\ket*{I^{(5/2)}}\cr
L_1\ket*{I^{(5/2)}}&=\qty( -b_{1/2}^2  +\frac{3}{2}{b}_{5/2}\pdv{b_{3/2}} +\frac{1}{2}{b}_{3/2}\pdv{b_{1/2}} )\ket*{I^{(5/2)}} \cr
L_0\ket*{I^{(5/2)}}&= \qty(\frac{1}{2} b_{1/2} \pdv{b_{1/2}}+\frac{3}{2} {b}_{3/2}\pdv{b_{3/2}}+\frac{5}{2} {b}_{5/2}\pdv{b_{5/2}} )\ket*{I^{(5/2)}} .
\end{aligned}
\ee
Similar to the rank-$\frac{3}{2}$ case, we perform the change of variables so that $\Lambda_k$ becomes the corresponding eigenvalue of $L_k$, that is
\be 
\Lambda_5= -b_{5/2}^2~, \qquad 
\Lambda_4= -2b_{3/2}b_{5/2}~, \qquad 
\Lambda_3= -b_{3/2}^2-2 b_{1/2} b_{5/2},
\ee 
and normalize the state as
\be
\ket*{I^{(5/2)}}  \longrightarrow \exp(\frac{47 \Lambda_4^5}{960 \Lambda_5^4}-\frac{5 \Lambda_3 \Lambda_4^3}{24 \Lambda_5^3}+\frac{\Lambda_3^2 \Lambda_4}{4 \Lambda_5{ }^2}) \ket*{I^{(5/2)}}.
\ee
Then, we obtain the following representation of Virasoro algebra
\be
\begin{aligned}
& L_5\ket*{I^{(5/2)}}=\Lambda_5\ket*{I^{(5/2)}}, \cr
& L_4\ket*{I^{(5/2)}}=\Lambda_4\ket*{I^{(5/2)}}, \cr
& L_3\ket*{I^{(5/2)}}=\Lambda_3\ket*{I^{(5/2)}}, \cr
& L_2\ket*{I^{(5/2)}}=\left(-\frac{\left(\Lambda_4\right)^3}{3\left(\Lambda_5\right)^2}+\frac{\Lambda_3 \Lambda_4}{\Lambda_5}+\Lambda_5 \frac{\partial}{\partial \Lambda_3}\right)\ket*{I^{(5/2)}}, \cr
& L_1\ket*{I^{(5/2)}}=\left(\frac{\left(\Lambda_4\right)^4}{3\left(\Lambda_5\right)^3}-\frac{\Lambda_3\left(\Lambda_4\right)^2}{\left(\Lambda_5\right)^2}+\frac{\left(\Lambda_3\right)^2}{\Lambda_5}+3 \Lambda_5 \frac{\partial}{\partial \Lambda_4}+2 \Lambda_4 \frac{\partial}{\partial \Lambda_3}\right)\ket*{I^{(5/2)}}, \cr
& L_0\ket*{I^{(5/2)}}=\left(5 \Lambda_5 \frac{\partial}{\partial \Lambda_5}+4 \Lambda_4 \frac{\partial}{\partial \Lambda_4}+3 \Lambda_3 \frac{\partial}{\partial \Lambda_3}\right)\ket*{I^{(5/2)}},
\end{aligned}
\ee
which is the same representation as given by the recursion relation \eqref{recursion}. This again proves the equivalence between the parameterization methods. For the generic $n$, one can check that by matching the eigenvalues of $\qty{L_k}_{k=n}^{2n-1}$, the change of variable should always produce the correct differential operators. The remaining coefficients can then be fixed by a proper normalization.

\acknowledgments
The authors are grateful to Ziheng Cao, B. Haghighat, T. Nishinaka, Yutai Zhang for valuable discussions and correspondence. 
This work is supported by the Shanghai Municipal Science and Technology Major Project (No. 24ZR1403900).

\bibliographystyle{JHEP}
\bibliography{references}

\end{document}